      [1999/12/01 v1.4c Il Nuovo Cimento]
\documentclass{cimento}
\usepackage{graphicx}

\title{$D_{sJ}(2700)$ and other puzzling states in charm spectroscopy}%
\author{S.~Nicotri\from{a}\from{b}}
\instlist{\inst{a} Universit\`a degli Studi di Bari,
Dipartimento di Fisica \inst{b}Istituto Nazionale di Fisica
Nucleare, Sezione di Bari, Italy}

\def \G {\Gamma}

\def \be {\begin{equation}}
\def \ee {\end{equation}}
\def \bea {\begin{eqnarray}}
\def \eea {\end{eqnarray}}
\def \non {\nonumber}

\def \ra {\rightarrow}

\def \pr {\prime}

\def\laq{~\raise 0.4ex\hbox{$<$}\kern -0.8em\lower 0.62
ex\hbox{$\sim$}~}
\def\gaq{~\raise 0.4ex\hbox{$>$}\kern -0.7em\lower 0.62
ex\hbox{$\sim$}~}

\begin{document}

\maketitle

\begin{abstract}
I briefly introduce the status of charm physics and I discuss
the case of the meson with open charm $D_{sJ}(2700)$.
\end{abstract}

\section{Introduction}

This period can be considered a sort of ``Renaissance'' for
charm physics. Recently, a great number of new states have been
discovered in both the open and the hidden charm sector. Among
these there are $D_{sJ}^*(2317)$, $D_{sJ}(2460)$,
$D_{sJ}(2700)$, $D_{sJ}(2860)$, $D_{0}^*(2308)$,
$D_{1}^\prime(2440)$, $h_c$, $\eta_c^\prime$, $X(3872)$,
$X(3940)$, $Y(3940)$, $Z(3930)$, $Y(4260)$, the mysterious
$Z(4430)$\ldots Part of the renewed interest growing around this
branch of particle physics is due to the fact that some of the
above states fail to fit into standard classification. This fact
has lead people to propose many interpretations, like molecular
charmonium, hybrids, orbital and radial excitations, tetraquarks,
and so on. In this paper I discuss the case of the $c\bar s$
meson $D_{sJ}(2700)$, giving hints on how to classify it. For a
more comprehensive discussion on charm physics the reader can
refer to \cite{review}.

\section{$D_{sJ}(2700)$}

$D_{sJ}(2700)$ has been observed by Belle Collaboration in the
Dalitz plot analyses of the process $B^+\to K^+D^0\bar D^0$ as
accumulation of events at $M^2(D^0 K^+) = 7-8$ GeV$^2$ in
the invariant mass distribution $M^2(D^0 K^+)$
\cite{Abe:2006xm}. The branching fractions ${\cal B}(B^+ \to
{\bar D}^0 D^0 K^+)=(22.2 \pm 2.2|_{stat} {^{+2.6}_{-2.4}}
|_{sys})\times10^{-4}$ and ${\cal B}(B^+ \to D_{sJ}(2700) {\bar
D}^0) \times {\cal B}(D_{sJ}(2700) \to D^0 K^+)=(11.3
\pm2.2|_{stat}{^{+1.4}_{-2.8}}|_{sys}) \times 10^{-4}$ have been
measured. Measured mass and width are respectively
$M=2708\pm9^{+11}_{-10}$~MeV, $\G=108\pm23^{+36}_{-31}$~MeV.
Moreover, from the distribution in the helicity angle $\theta$,
the angle between the $D^0$ momentum and the opposite of the
kaon momentum in the $D^0\bar D^0$ rest frame, it was possible
to assign spin-parity $J^P=1^-$ to $D_{sJ}(2700)$.

The classification of the $c\bar s$ states is easier in the
heavy-quark limit $m_c\ra\infty$. In this limit the spin $s_Q$
of the heavy quark and the total angular momentum $s_\ell$ of
the meson light degrees of freedom: $s_\ell=s_{\bar q}+ \ell$
($s_{\bar q}$ light antiquark spin, $\ell$  orbital angular
momentum of the light degrees of freedom relative to the heavy
quark) are decoupled, and the spin-parity  $s_\ell^P$  is
conserved in processes involving strong interactions
\cite{HQET}. This allows to classify mesons into doublets
labeled by $s_\ell^P$, each containing a pair of mesons with
$J^P=(s_\ell^P-1/2,s_\ell^P+1/2)$ and degenerate in mass. The
standard classification of known $c\bar s$ states in this scheme
is given in Table~\ref{organiz} \cite{PDG}.
\begin{table}[ht]
\begin{tabular}{c|cccccc}
\hline $s_\ell^P  $ & ${1\over 2}^-$ & ${1\over 2}^+$ &
${3\over2}^+$ & ${3\over 2}^-$ & ${5\over 2}^-$ \\ \hline
$J^P=s_\ell^P-{1\over 2}$ & $D_s (1965) \,\, (0^-)  $ & $D^*_{sJ}(2317) \,\, (0^+)$ & $D_{s1}(2536) \,\, (1^+) $ & $(D^{*}_{s1})  \,\, (1^-)$ & $(D^{*\prime}_{s2}) \,\,(2^-)$\\
$J^P=s_\ell^P+{1\over 2}$ & $D_s^*(2112) \,\, (1^-)$ &
$D_{sJ}(2460) \,\, (1^+)$  & $D_{s2}(2573) \,\,
(2^+)  $ & $(D^{*}_{s2}) \,\, (2^-)$ & $ (D_{s3}) \,\, (3^-)$\\
\hline
\end{tabular}
\caption{$c \bar s$ states organized according to $s_\ell^P$ and
$J^P$. The mass of known mesons is indicated.}\label{organiz}
\end{table}
The states labeled by $D^{*}_{s1}$, $D^{*\prime}_{s2}$,
$D^{*}_{s2}$ and $D_{s3}$ are still to be discovered. There are
hints of the fact that the $D_{s3}$ state is the
$D_{sJ}(2860)$ meson \cite{dsj2860,vanbeveren}. In this picture
$D_{sJ}(2700)$ could be the $D^{*}_{s1}$ state of the
$s_\ell^P=3/2^-$ doublet or $D^{*\pr}_s$, the first radial
excitation of $D_s^*(2112)$. A way to select the right
assignment between these two is to examine ratios of partial
decay widths. Using an effective Lagrangian describing strong
decays of heavy mesons to final states comprising a light
pseudoscalar meson, and displaying heavy quark and chiral
symmetries, the ratios $R_1=\frac{\G(D_{sJ}\ra
D^*K)}{\G(D_{sJ}\ra DK)}$, $R_2=\frac{\G(D_{sJ}\ra
D_s\eta)}{\G(D_{sJ}\ra DK)}$ and $R_3=\frac{\G(D_{sJ}\ra
D^*_s\eta)}{\G(D_{sJ}\ra DK)}$ have been calculated with results
given in Table \ref{results} \cite{dsj2700}.
\begin{table}[h]
      \begin{center}
        \begin{tabular}{c|ccc}
          \hline
          & $R_1 \times 10^{2} $ & $R_2 \times 10^{2} $ & $R_3 \times 10^{2}$ \\
          \hline
          $D_s^{*\prime}$ & {$91 \pm 4$} & $20 \pm 1$ & {$5 \pm 2$}  \\
          \hline
          $D_{s1}^*$ &{$4.3 \pm 0.2$} & $16.3 \pm  0.9$ & {$0.18 \pm 0.07 $} \\
          \hline
        \end{tabular}
      \end{center}
      \caption{Ratios $R_i$ for $D_{sJ}(2700)$ identified as
      $D_s^{*\prime}$ ($ s_\ell^P={1 \over 2}^-$) or
       $D_{s1}^*$
      ($s_\ell^P={3 \over 2}^-$).}\label{results}
    \end{table}
The ratios $R_1$ and $R_3$ are very different if $D_{sJ}(2700)$
is $D^*_{s1}$ or $D^{*\pr}_s$, so that the measurements of these
ratios allow to properly identify the $D_{sJ}(2700)$. In
particular the decay mode to $D^*K$ has very different branching
ratios in the two possible assignments, so that a measurement of
such a branching fraction would be useful to identify
$D_{sJ}(2700)$. Within the same framework, individual branching
ratios have also been calculated and they are shown in Table
\ref{br-s} \cite{dsj2700}.

\begin{table}
\begin{center}
\begin{tabular}{c|ccc}
  \hline
     & ${\cal B}(D_{sJ} \to D^0K^+)$ & ${\cal B}(D_{sJ} \to D^+K_S)$ & ${\cal B}(D_{sJ} \to D_s\eta)$ \\
  \hline
  $D_s^{*\prime}$ & $(24 \pm 14 )\%$ & $(12 \pm 7.0)\%$ & $( 7\pm 4 )\%$ \\
  \hline
  $D_{s1}^*$ & $( 44 \pm 25 )\%$ & $ (21\pm 12 )\%$ & $(11\pm 6)\%$ \\
  \hline
\end{tabular}
\end{center}
\begin{center}
\begin{tabular}{c|ccc}
   \hline
     & ${\cal B}(D_{sJ} \to D^{*0}K^+)$ & ${\cal B}(D_{sJ} \to D^{*+}K_S)$ & ${\cal B}(D_{sJ} \to D_s^*\eta)$ \\
   \hline
   $D_s^{*\prime}$ & $( 22\pm  13)\%$ & $( 10\pm 6 )\%$ & $( 1.7\pm 1.2)\%$\\
   \hline
   $D_{s1}^*$ & $( 1.9\pm  1.1)\%$ & $ (0.9\pm 0.5 )\%$ & $(0.12\pm 0.09 )\%$\\
   \hline
\end{tabular}
\end{center}
\caption{$D_{sJ}(2700)$ branching fractions corresponding to the
two assignments.} \label{br-s}
\end{table}

Since in the heavy quark limit the heavy mesons are collected
in doublets with a definite value of $s_\ell^P$, the state
$D_{sJ}(2700)$ has a partner  from which it differs only for the
value of the total spin.

The partner of $D_s^{*\prime}$ ($s_\ell^P=1/2^-$) has $J^P=0^-$;
it is denoted $D_s^\prime$, the first radial excitation of
$D_s$. On the other hand, the partner of $D_{s1}^*$
($s_\ell^P=3/2^-$) has $J^P=2^-$ ($D_{s2}^*$). In both cases,
the decay modes $D_s^\prime$, $D_{s2}^*$ $\to D^{*0} K^+$, $
D^{*+} K^0_{S(L)}$, $ D^{*}_s \eta$, are permitted. In the heavy
quark limit, these partners are degenerate, hence, assigning
them the same mass as $D_{sJ}(2700)$ one finds:
\begin{eqnarray}
\Gamma(D_s^\prime)&=& (70 \pm 30) \,\,{\rm MeV} \non \\
\Gamma(D_{s2}^*)&=& (12 \pm 5) \,\,{\rm MeV}\non
\label{gammatotspinpartners}
\end{eqnarray}
and the  branching fractions in Table \ref{spin-partner}.
Therefore, in the two assignments the spin partners differ for
their decay width.
\begin{table}[h]
\begin{center}
\begin{tabular}{c|c | c| c} \hline
  & ${\cal B} (D_s^{\prime}(D_{s2}^*)\to D^{*0}K^+)$ & ${\cal B}(D_s^{\prime}(D_{s2}^*)\to D^{*+}K_S)$ &
     ${\cal B}(D_s^{\prime}(D_{s2}^*)\to D_s^*\eta)$
     \\\hline
  $D_s^{\prime} \, (J^P=0^-)$ & $( 50.0\pm  0.5)\%$ & $( 23.7\pm 0.2 )\%$ & $( 2.6\pm 0.9)\%$  \\ \hline
  $D_{s2}^* \, (J^P=2^-)$ & $( 49.8\pm  0.6)\%$ & $ (23.6\pm 0.2 )\%$ & $(3.1\pm 1.0 )\%$ \\
  \hline
\end{tabular}
\end{center}
\caption{Branching ratios of  the spin partner of  $D_{sJ}(2700)$ for
the two quantum number assignments.} \label{spin-partner}
\end{table}

\acknowledgments I thank P.~Colangelo, F.~De~Fazio and
F.~Giannuzzi for collaboration.

\end{document}